\documentclass[preprint,12pt]{elsarticle}
\usepackage{xcolor}
\usepackage[hypertex]{hyperref}

\usepackage{graphicx,subfigure}
\usepackage{amsmath,amsfonts,amssymb}
\graphicspath{{figures/}}

\newcommand{\f}{\frac}
\newcommand{\suml}{\sum\limits}
\newcommand{\m}{\mathbf}

\newcommand{\bs}{\boldsymbol}
\graphicspath{{figures/}}

\journal{Physica A}

\begin{document}

\begin{frontmatter}



\title{On the relation between Vicsek and Kuramoto models of spontaneous synchronization}


\author{A.A. Chepizhko}
\author{V.L. Kulinskii}
\address{Department of Theoretical Physics, Odessa National University, Dvoryanskaya 2, 65026 Odessa, Ukraine}

\begin{abstract}
The Vicsek model for the self-propelling particles in 2D is investigated with respect to addition of the stochastic perturbation of dynamic equations. We show that this model represents in essential the same type of bifurcations under different type of noise as the celebrated Kuramoto model of spontaneous synchronization. These models demonstrate similar behavior at least within the mean-field approach. To prove this we consider two types of noise for the Vicsek model which commonly considered in the literature: the intrinsic and the extrinsic ones (according to the terminology of J.~A.~Pimentel \textit{et al.} [Phys. Rev. E {\bf 77}, 061138 (2008)]). The qualitative correspondence with the bifurcation of stationary states in Kuramoto model is stated. New type of stochastic perturbation - the ``mixed`` noise is proposed. It is constructed as the weighted superposition of the intrinsic and the extrinsic noises. The corresponding phase diagram ``noise amplitude vs. interaction strength`` is obtained. The possibility of the tricritical behavior for the Vicsek model is predicted.
\end{abstract}

\end{frontmatter}

\section{Introduction}
The Vicsek model (VM) of the dynamics of self-propelling particles proposed in \cite{spp_cva_prl1995}
gave intense impact for the research of the order-disorder
transitions in the systems of such kind. In this model the
motion of self-propelling particles is driven by very simple
rules. The dynamic rule proposed by Vicsek is based on the constraint of alinement of particle velocity along the direction of the mean velocity of its neighborhood. As shown by numerous simulations \cite{spp_csahokczirok_pha1997,spp_gregchate_prl2004} such dynamic rule favors the emergence of ordered motion at high enough densities.

Addition of the
stochastic perturbation in the direction of motion leads
to the transition from the ordered state of synchronized motion to the disordered state when the intensity of the
perturbation grows.
As it has been demonstrated in
\cite{spp_aldanahuepe_prl2007,spp_kineticus_proc2009}, the
character of such transition depends essentially on the type of the stochastic perturbation. The intrinsic noise which is nothing but adding the stochastic increment to the direction of particle's motion originally used in \cite{spp_cva_prl1995},
leads to the continuous phase transition. The so called extrinsic noise introduced in \cite{spp_gregchate_prl2004} where some
random vector is added to the particle velocity, makes the
transition discontinuous. It should be noted that the same
situation is well known for another model of synchronization - the Kuramoto model (KM) \cite{spp_kuramoto1975} (see also
\cite{spp_kuramotomodel_rmp2005} and references therein).
Yet in the KM the type of the transition depends on the distribution function $g(\omega)$ of the own frequencies of the oscillators \cite{spp_kuramotomodel_rmp2005}.

In this paper we investigate the relation between the
Viscek and the Kuramoto models.
Also we extend the results of \cite{spp_kineticus_proc2009}
by introducing the novel type of noise perturbation which is the weighted sum of the intrinsic and the extrinsic ones. We call it ``mixed`` noise. It is shown that for
such noise the bifurcation of stable solution becomes more
complex. The tricritical behavior
similar to that in Kuramoto model \cite{spp_kuramototricrit_physd1998}
is demonstrated.

\section{Relation between Vicsek and Kuramoto models}
\label{sec_vicsekutamoto}
The dynamic equation in the continuous time limit for the Vicsek model was considered in \cite{spp_kineticus_proc2009,spp_usdiscr_physa2007}.
It has a simple form:
\begin{equation}\label{eq:eq_eom}
\frac{d}{dt}\m{v}_{i}=\bs{\omega}_{\m{v}_i}\times \m{v}_{i}\,\,,
\end{equation}
where $\bs{\omega}_{\m{v}_i}$ is the ``angular velocity`` of $i$-th particle. This angular velocity depends on the velocities of neighboring particles. The self-propelling force and the frictional force are assumed to balance each other. As it is shown in
\cite{spp_kineticus_proc2009} the angular velocity
$\bs{\omega}_{\m{v}_{i}}$ of the dynamic model,
which corresponds to the discrete Vicsek automaton in the continuous time limit, consists of two parts:
\begin{equation}\label{eq:om}
  \bs{\omega}_{\m{v}_{i}} = \bs{\omega}_{\m{u}_{i}} +
  \bs{\omega}_{\m{v}\m{u}}\,,
\end{equation}
where $\bs{\omega}_{\m{u}_{i}}$
is the angular velocity corresponding to the average velocity of the ``nearest`` (in the sense of some model for the distance) neighbors of $i$-th particle:
\begin{equation}\label{eq:orderparameter}
   \m{u}_i=  \f{1}{N_i}\suml_{
\left\langle\, i,j \,\right\rangle
   }\,\m{v}_j\,,
\end{equation}
which serves as the local order parameter. Here $N_i$ is the number of the nearest neighbors of $i$-th particle.
Another term:
\begin{equation}\label{eq:omega_uv}
\bs{\omega}_{\m{v}\m{u}} =\, A\,\m{v}_{i}\times \bs{\tau}_{\m{u}_{i}}\,
\end{equation}
denotes the relative angular velocity of the particle
with respect to the direction $\bs{\tau}_{\m{u}_{i}} = \f{\m{u}_{i}}{\left|\m{u}_{i}\right|}$ of the average
velocity of the neighbors \eqref{eq:orderparameter}.
The coefficient $A$ in Eq. ~\eqref{eq:omega_uv} determines the relaxation rate (inverse of characteristic relaxation time) to the direction of the
local order parameter \cite{spp_kineticus_proc2009}. Such rate
depends both on the number of the nearest neighbors and the value of the average velocity.
 Therefore it is natural
that $A$ is proportional to the local flux:
\begin{equation}\label{eq:a_model}
A = \lambda \,u_i\,,
\end{equation}
where $\lambda $ is proportional to the number of the neighbors. As it is shown in \cite{spp_kineticus_proc2009} such dependence can be obtained in the approximation when the velocity and the density fields are decoupled.
Throughout this paper we consider 2D case and use the following definition of the local order parameter:
\begin{equation}\label{eq:km_orderparameter}
  r_i\,e^{i\,\psi_i} =\f{1}{N_i} \suml_{
\left\langle\, i,j \,\right\rangle
  } e^{i\,\theta_{j}}\,.
\end{equation}
In terms of this order parameter the equation
\eqref{eq:om} takes the form \cite{spp_kineticus_proc2009}:
\begin{equation}\label{eq:vkm}
  \dot{\theta}_{i} =\dot{\psi_{i}} +A\,\sin\left(\,\psi_i - \theta_i \,\right)\,,
\end{equation}
where $A$ is given by \eqref{eq:a_model}.
%

The dynamic equation similar to \eqref{eq:vkm} appears in the theory of stochastic synchronization for the KM \cite{spp_kuramoto1975}.
The KM is determined by the following dynamic equations for the oscillators phases:
\begin{equation}\label{eq:kmlocal}
  \dot{\theta}_{i} =\omega_{i} +
  K\,\suml_{\left\langle\,i,j\,\right\rangle }\sin\left(\,\theta_j - \theta_i
  \,\right) = \omega_{i}  + K\,N_i\,r_i\,\sin\left(\,\psi_i - \theta_i
  \,\right),
\end{equation}
where $K$ - is the interaction strength, $\psi_i$ is
the average local phase of the nearest oscillators \cite{spp_kuramotomodel_rmp2005}.

From the mathematical point of view dynamic equations of motion Eq.~\eqref{eq:kmlocal} and Eq.~\eqref{eq:vkm} are  similar.
The main difference between \eqref{eq:vkm} and \eqref{eq:kmlocal} is the form of the first term. For the Kuramoto model it defines the own frequency of the $i$-th oscillator. Usually the distribution of own frequencies is modeled by some function $g(\omega)$. For the Vicsek model the first term determines the collective contribution which is assumed to vary slowly as it is a collective mode. Clearly, it vanishes in the case of coherent motion, yet this term favors the transition to the ordered state. From this it follows that the synchronization in the VM is more pronounced than that in the KM with finite radius of interaction. Note that this term can be connected with the convective term $\m{v}\cdot\nabla \m{v}$ in the continuum approach. According to the arguments in \cite{spp_tonertu_prl1995} the convective term effectively generates the long-ranged correlations. In its turn this leads to the emergence of the ordered motion.

Based on the mathematical similarity of the dynamic equations for the VM and the KM mentioned above, it is quite natural to expect that these models demonstrate qualitatively the same behavior at least in the mean field approximation. Indeed, the VM with intrinsic noise shows the supercritical bifurcation from disordered to ordered state \cite{spp_cvstanley_jphysa1997}. The same behavior is observed for the KM with unimodal frequency distribution of the form $g(\omega) = \delta(\omega)$ \cite{spp_strogartz_prl1992}. Yet in the KM the type of the transition depends on the sign of $g''(0)$ \cite{spp_kuramotomodel_rmp2005}. In the framework of the VM the distribution of the own frequencies is equivalent to the introduction of the stochastic perturbation into the equation of motion. Therefore for the VM there must be the dependence of the character of the order-disorder transition on the type of the stochastic perturbation. This result was obtained in numerical simulations in \cite{spp_gregchate_prl2004} and confirmed theoretically \cite{spp_kineticus_proc2009,spp_aldananoiseswarming_pre2008}.

\begin{figure}
	\centering
		\includegraphics[scale=0.5]{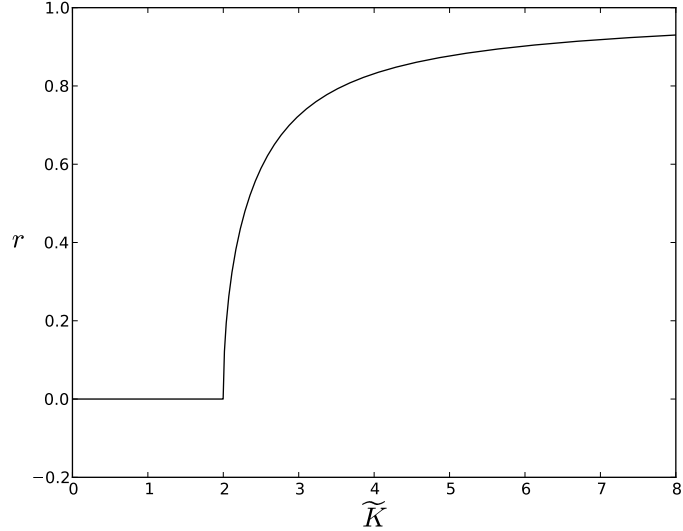}
	\caption{The solution of Eq.~\eqref{eq:order2}. $\widetilde{K}= \frac{K}{D}$. Supercritical behavior of order parameter is shown. It is similar to the one in Vicsek model with intrinsic noise.}
	\label{fig:kuramoto_solution}
\end{figure}

First let us consider the case of intrinsic noise \cite{spp_cva_prl1995} which can be modeled by inclusion of the Langevin source into the equation of motion with
$g(\omega )= \delta (\omega )$:
\begin{equation}\label{eq:langeven_scalar}
d{\theta_{i}}=-K\,r\sin(\theta_{i})\,dt+\sqrt{D} dw_i(t)\,.
\end{equation}
Here $w_i(t)$ is the Wiener process \cite{book_vankampen} and represents the random increment of the angle and $D$ is the diffusion coefficient. In such an approach it represents the intensity of the intrinsic noise in the particle decision algorithm \cite{spp_aldananoiseswarming_pre2008}. In the mean-field approximation one easily comes to the
self-consistent equation for the stationary value
of the order parameter \cite{spp_kineticus_proc2009}:
\begin{equation}\label{eq:order2}
r=\frac{I_{1}(\frac{Kr}{D})}{I_{0}(\frac{Kr}{D})} = F(Kr)\,,
\end{equation}
where $I_0$ and $I_1$ are modified Bessel functions of the first kind \cite{book_abramovitzstegun}.
This equation can be easily solved numerically (see Fig.~\ref{fig:kuramoto_solution}).
The nontrivial solution appears continuously from the trivial one.

As the second case we consider the extrinsic noise perturbation for the VM introduced in \cite{spp_gregchate_prl2004}. This perturbation corresponds to the stochastic deviation of the direction of motion of $i$-th particle due to addition of the random vector $\bs{\xi}_i$ with the amplitude $|\bs{\xi}_i| = \xi\,N_{i}$:
\begin{equation}\label{extrinsicnoise}
    \theta_i (n + 1) = \rm{angle} (\bs{u}_i (n) + \bs{\xi}_i)\,,
\end{equation}
where $\xi$ is the control parameter. Let us define the quantity $K = 1 / \xi$ then the the self-consistent mean-field equation for the order parameter is as follows \cite{spp_kineticus_proc2009}:
\begin{equation}
r = \frac{1}{2 \pi} \int \limits_{-\pi}^{\pi} \frac
{\left(\,K r + \cos \alpha\,\right) d \alpha}
{\sqrt{1 + 2 Kr \cos \alpha + \left(\,Kr\,\right)^2}} = F (K r) \,.
\label{eq:self2}
\end{equation}
It is quite remarkable that \eqref{eq:self2} coincides exactly with corresponding equation for the order parameter of the Kuramoto-like model augmented with the phase pinning \cite{spp_strogatzpinning_prl1988}.

\begin{figure}[hbt!]
	\centering
		\includegraphics[scale=0.4]{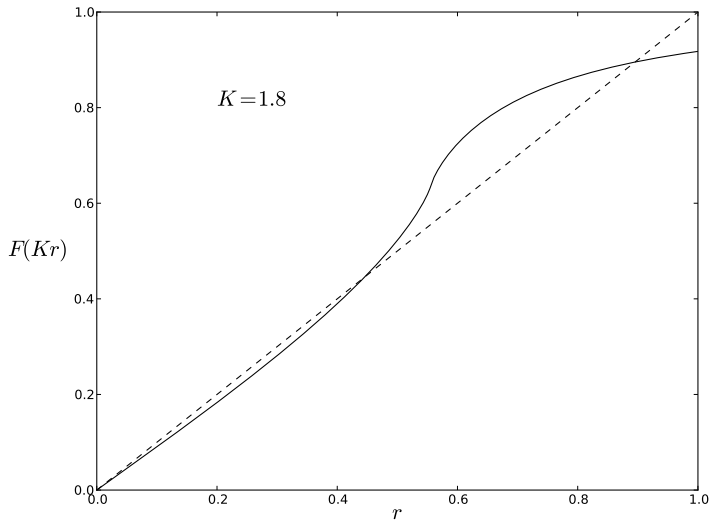}
	\caption{Graphical solution of Eq. \eqref{eq:self2}.}
	\label{fig:K_Str_bw}
\end{figure}

In this case the subcritical bifurcation of the solution occurs (see Fig.~\ref{fig:K_Str_bw}). There is a discontinuous jump between ordered and disordered motion because of the existence of the metastable state.
Therefore the extrinsic noise in the VM is equivalent to the
random pinning for the direction of motion. For the
self-propelling particle this perturbation corresponds to the
instant change of the direction due to triggering the attention to some particle which moves asynchronously with the nearest
neighbors forming a stable flock. Due to the nonlinear
character of the angular dependence this ``extrinsic``
perturbation corresponds to the ``colored`` Langevin force and indeed represents physical situation quite different from that for the intrinsic noise  \cite{spp_aldanahuepe_prl2007,spp_aldananoiseswarming_pre2008}.

\section{``Mixed`` noise}\label{sec_mixednoise}
The similarity between the bifurcations of the stationary states in the Vicsek and the Kuramoto
models is based on the equivalence of their equations of motion. The fact that the Vicsek model with the extrinsic noise
demonstrates the hysteretic behavior typical for the first
order phase transition and continuum phase transition for the
intrinsic noise, allows us to conclude that these perturbations can be present simultaneously in the general case of the dynamics. Therefore
the tricritical behavior might be expected if some switching parameter between these two perturbations is added. Note that tricritical behavior is known for the Kuramoto model with symmetrical bimodal distribution function
\begin{equation}\label{kuramotobimodal_mixing}
g_{\eta}(\omega) = \eta \delta (\omega -\omega_0)+(1-\eta)\delta (\omega +\omega_0),
\end{equation}
with $\eta = 1/2$ (see \cite{spp_kuramototricrit_physd1998}).

The simplest variant is the statistical mixing similar to that of \eqref{kuramotobimodal_mixing} the extrinsic and the intrinsic noises. The corresponding  self-consistent equation is of the form:
\begin{equation}\label{eq:mix_self_c}
r = \eta\, F_{v}(r)+ (1-\eta)\,F_{s}(r)\,.
\end{equation}
Here the right hand side of \eqref{eq:mix_self_c} corresponds to the statistically mixed probability density for the angle of direction:
\begin{equation}
f_{mix} (\theta,r) = \eta f_{v} (\theta,r) + (1-\eta) f_{s} (\theta, r)\,,
\label{eq:mixed_distr}
\end{equation}
where $f_{s}$ and $f_{v}$ are distribution functions for intrinsic and extrinsic noises respectively and $0\le \eta \le 1$ denotes the parameter of statistical mixing.

\begin{figure}
	\centering
		\includegraphics[scale = 0.4]{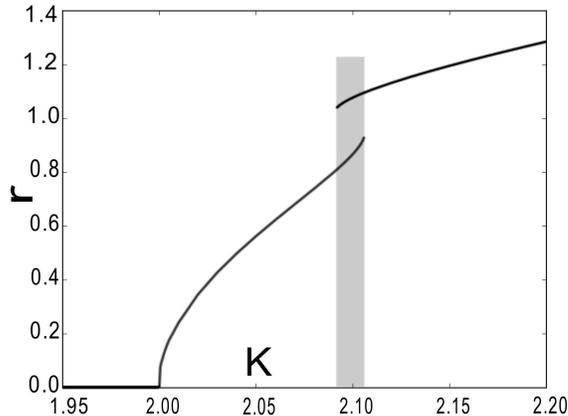}
	\caption{Solutions for the order parameter $r$ vs. $K$. Gray region shows the situation when two stable regimes are possible. The value of mixing parameter $\eta \approx 0.16$.}
	\label{fig_mixingtricrit}
\end{figure}

As the analysis of \eqref{eq:mix_self_c} shows, the bifurcation of its solutions with varying parameter $\eta$ corresponding to the tricritical behavior (see Fig.~\ref{fig_mixingtricrit}). Note that the same behavior is obtained for the KM if the number of modes included into frequency distribution function $g(\omega)$ more than one (see \cite{spp_kuramototricrit_physd1998,spp_kuramotosynchro_jsp1992,
spp_kuramototrimodal_pre2001}\label{citationreviewer}).

Considering \eqref{eq:mixed_distr} as the stationary solution of the Fokker-Plank equation and using the standard relation with the Langevin equation (see e.g. \cite{book_vankampen}):
\begin{equation}\label{eq:mixed_L}
d \theta_i = -K\,r\sin(\theta_{i})\,dt + b (\theta_i, r) dw_i(t)
\end{equation}
one can show that:
\begin{equation}
b^2(\theta,r; \eta) = \frac{1}{f_{mix}(\theta,r)}\,\int \limits_{\theta_0}^{\theta} K\, r \,f_{mix}(x,r)\,\sin{x} \, d \,x\,,
\label{eq:B_x}
\end{equation}
where $\theta_0$ can be found from the condition that in accordance with \eqref{eq:langeven_scalar} for the intrinsic noise ($\eta = 0$) $b(\theta,r) = 1$ :
\[
\theta_0 \equiv \arccos \left(\frac{\ln 2}{Kr} + \cos \theta \right)\,.
\]
Then from \eqref{eq:B_x} one can easily get the following expression for $b(\theta,r;\eta)$:
\begin{equation}
b^2 (\theta, r; \eta)= (1 - \eta) + \eta  (B_{ext}(\theta) - B_{ext}(\theta_0))\,,
\label{eq:b_integr}
\end{equation}
where the function:
\[
B_{ext} (\theta) \equiv \frac{2 r^2}{\sqrt{1 + r^2}}\arctan\left(\frac{\sqrt{1 + r^2 + 2 r \cos \theta}}{\sqrt{1+r^2}}\right) - \sqrt{r^2 +2 r \cos \theta + 1}\,,
\]
determines the corresponding parameter
of Fokker-Plank equation in case of extrinsic noise ($\eta = 1$).
\begin{figure}
	\centering
		\includegraphics[scale = 0.4]{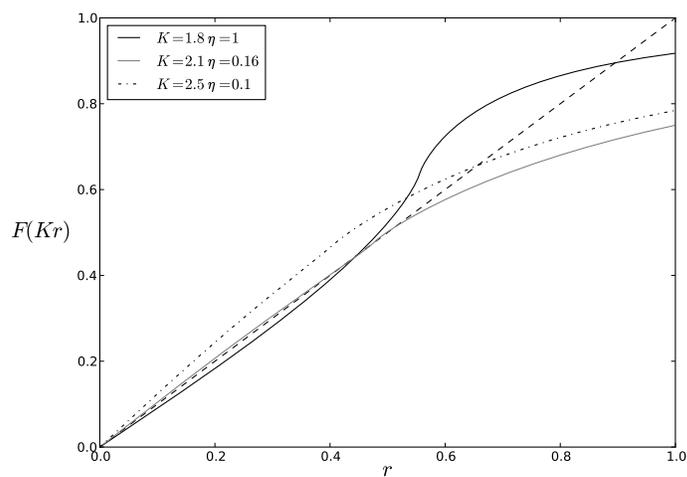}
	\caption{Graphical solution of self-consistent equation for the case of mixed noise with different values of parameters $\eta$ and $K$}
	\label{fig:diff-beh}
\end{figure}
\begin{figure}
	\centering
		\includegraphics[scale=0.37]{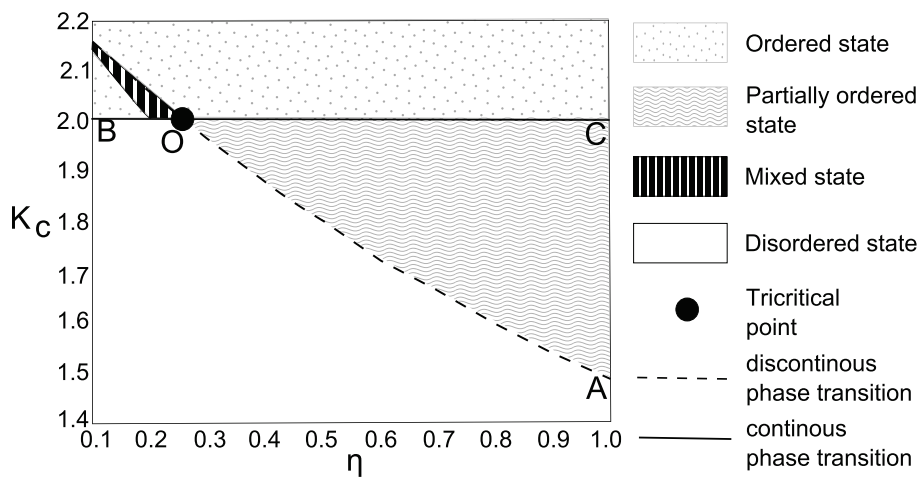}
	\caption{Phase diagram for ``mixed`` noise.
	}
\label{fig:phase_diagr}
\end{figure}

As the analysis shows, the addition of ``mixed`` noise leads to complex behavior. It is natural that for $\eta \to 1$ the behavior corresponds to one with intrinsic noise where the supercritical bifurcation takes place. Small values of mixing parameter $\eta$ drives the system into the situation with intrinsic noise for which the subcritical regime occurs. By appropriate tuning the parameter $\eta$ the possibility when both behaviors supercritical and subcritical exist at one time can be realized. Indeed for the case of ``mixed`` noise there is a region of values $K$ and $\eta$ where two nontrivial stable regimes exist simultaneously (see Fig.~\ref{fig_mixingtricrit}). The complete phase diagram for the mixed noise is shown in Fig.~\ref{fig:phase_diagr}.  Line $AO$ stands for the discontinuous phase transition (like in extrinsic noise case). Line $BC$ is for the continuous phase transition (like in case of the intrinsic noise). The point $O$ is the tricritical point, when lines of phase transitions of different orders meet $(K = 2.0, \ \eta \approx 0.28)$.
Thus the statistical mixing of the extrinsic and the intrinsic noises leads to the emergence of the tricritical point. This prediction can be tested in computer simulation of the corresponding dynamics.

%

Note that the function $g(\omega)$ plays a
role of initial distribution for the frequencies and represents the background noise in the KM. In the Vicsek model there is no direct analog of such function. Yet as has been shown above, there is a correspondence between the bifurcation behavior of the KM determined by $g(\omega)$ (unimodal, bimodal, etc.) and the VM with specific kind of noise. It is interesting to find the relation between the properties of $g(\omega)$ and the type of noise in the VM which leads to the similarity of the phase diagram. The question whether the similarity between these models conserves with the inclusion of fluctuations, remains open.


\section{Discussion}
The results of the work can be summarized as follows. It is shown that the type of ordering for the Vicsek model depends on the kind of noise perturbation of the equation of motion. The transition to ordered state can be either of continuous or discontinuous for intrinsic and extrinsic noises correspondingly. 
It is demonstrated that the Vicsek model shows qualitatively the same variety of the phase behavior as the Kuramoto model of the phase synchronization of nonlinearly coupled oscillators. Both known cases of extrinsic and intrinsic noise action in Vicsek model demonstrate the results similar to those for Kuramoto model. In particular on the basis of the stated equivalence of these models  ``mixed`` noise is investigated and the tricritical behavior for the Vicsek model is demonstrated.

\section*{Acknowledgements}
The author V.~K. thanks to Prof. Signe Kjelstrup and
Prof. Dick Bedeaux for the kind invitation to the seminar in NTNU where the results have been discussed. The author A.~C. appreciates the financial support from the Organizers of Young Scientists Summer School in Statistical Physics, Lviv 2010.
The authors thank Dr. V. Ratushnaya for careful reading
of the manuscript and comments.
\newpage
%

%
\end{document}